\documentclass[useAMS,usenatbib]{mn2e}
\usepackage{amsmath}
\usepackage{epsfig}
\usepackage{txfonts}

\usepackage[pdftex,pdfpagemode={UseOutlines},bookmarks,bookmarksopen,colorlinks,linkcolor={blue},citecolor={blue},urlcolor={red}]{hyperref}
\usepackage{float}
\usepackage{graphicx}
\usepackage{caption}
\usepackage{subcaption}
\usepackage{natbib}
\usepackage{rotating,longtable,lscape}

\title[K-L cycles towards circumbinary planets]{Kozai-Lidov cycles towards the limit of circumbinary planets}

\author[Martin and Triaud]
{\parbox{\textwidth}{David V. Martin$^1$\thanks{E-mail: david.martin@unige.ch} and Amaury H.M.J. Triaud$^{2,3}$}
\vspace{0.4cm}\\
\parbox{\textwidth}{$^{1}$Observatoire de Gen\`eve, Universit\'e de Gen\`eve, 51 chemin des Maillettes, Sauverny 1290, Switzerland\\
$^{2}$ Centre for Planetary Sciences, University of Toronto at Scarborough, 1265 Military Trail, Toronto, ON, M1C 1A4, Canada\\
$^{3}$ Department of Astronomy \& Astrophysics, University of Toronto, Toronto, ON, M5S 3H4, Canada\\}}

\begin{document}

\date{Accepted 2015 September 21.  Received 2015 September 17; in original form 2015 August 25}

\pagerange{\pageref{firstpage}--\pageref{lastpage}} \pubyear{2015}

\maketitle

\label{firstpage}

\begin{abstract}

In this paper we answer a simple question: can a misaligned circumbinary planet induce Kozai-Lidov cycles on an inner stellar binary? We use known analytic equations to analyse the behaviour of the Kozai-Lidov effect as the outer mass is made small. We demonstrate a significant departure from the traditional symmetry, critical angles and amplitude of the effect. Aside from massive planets on near-polar orbits, circumbinary planetary systems are devoid of Kozai-Lidov cycles. This has positive implications for the existence of highly misaligned circumbinary planets: an observationally unexplored and theoretically important parameter space.

%We find that this is generally not possible, with the exception of some planets that are very massive, misaligned by close to $90^{\circ}$ and on very long-period orbits. We conclude that circumbinary planets may have stable, Kozai-free orbits at almost any mutual inclination. This has positive consequences for the existence of highly misaligned circumbinary planets: an observationally unexplored parameter space.
\end{abstract}


\section{Introduction}
\label{sec:intro}

\begin{figure*}  
\begin{center}  
\includegraphics[width=0.84\textwidth]{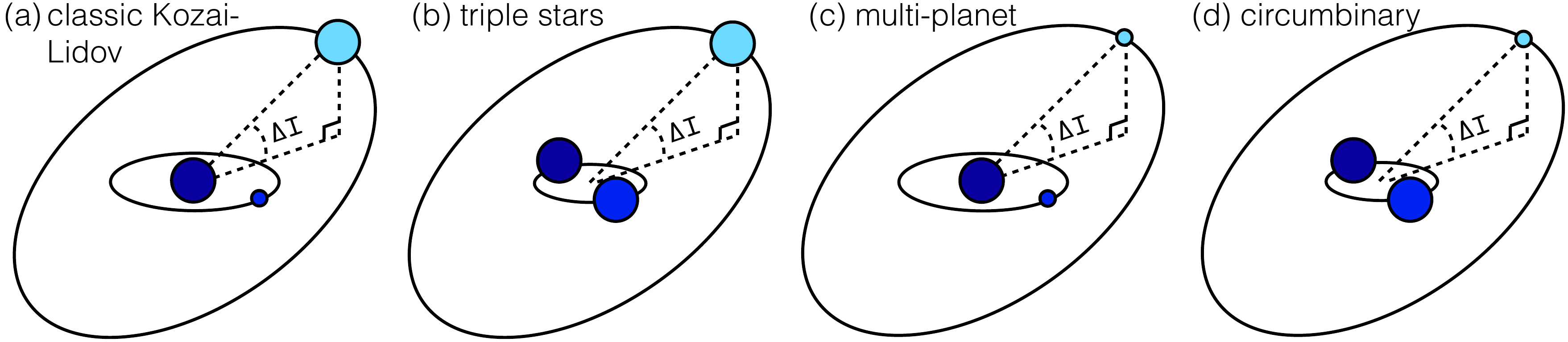} 
\caption{Zoo of different three-body configurations, all with orbits misaligned by $\Delta I$. The orbits are drawn with respect to the centre of mass of the inner orbit. Case (d) is qualitatively different to cases (a) - (c) because in (d) the outer orbit no-longer dominates the total angular momentum of the system.}
\label{fig:zoo_diagram}  
\end{center}  
\end{figure*}

In 1962 there were two independent papers published on an effect seen in the gravitational three-body problem. \citet{lidov62} investigated the effect the Moon has on artificial satellites orbiting the Earth. \citet{kozai62} looked at asteroids orbiting the Sun under the influence of Jupiter. Both systems were qualitatively the same: an inner restricted three-body problem with initially circular orbits and all of the orbital angular momentum confined to the outer orbit. (Fig.~\ref{fig:zoo_diagram}a). It was discovered that when the two orbital planes were significantly misaligned, between $39^{\circ}$ and $141^{\circ}$, there was a high-amplitude modulation in the eccentricity of the inner orbit (in their examples the satellite and the asteroid) and the mutual inclination. Contrastingly, the outer eccentricity was seen to remain constant. Overall, this is known as a Kozai-Lidov (K-L) cycle or effect.

The K-L effect was later generalised to the case of three bodies of comparable mass \citep{harrington68,lidov76} for the application to triple star systems (Fig.~\ref{fig:zoo_diagram}b). It has subsequently been implicated, for example, in the production of tight stellar binaries (period $\lesssim 7$ d). In this scenario the K-L cycle induced on the inner binary leads to close encounters between the two stars, at which point tidal friction dissipates orbital energy and shrinks the inner orbit (e.g. \citealt{mazeh79,eggleton06,fabrycky07}). 

The theory has been extended to eccentric outer orbits, in which case higher-order effects cause the system to be chaotic, possibly inducing flips in the inner orbit orientation. This has been applied to multi-planet systems (Fig.~\ref{fig:zoo_diagram}c) by \citet{naoz11} in order to explain why a surprisingly large fraction of hot-Jupiters are misaligned or even retrograde \citep{hebrard08,schlaufman10,triaud10}.

The similarity between the three scenarios in Fig.~\ref{fig:zoo_diagram}a - c is that the system's orbital angular momentum is largely confined to the outer orbit. The timescale, amplitude and limits of K-L cycles do vary between the different configurations, but for initially circular orbits there is qualitatively not a radical departure from the original work of \citet{lidov62} and \citet{kozai62}.

In this paper we explore a different type of three-body system: circumbinary planets (Fig.~\ref{fig:zoo_diagram}d). The discovery of eight transiting systems by the {\it Kepler} mission \citep{doyle11,welsh14} has given this class of planet widespread scientific exposure and validity. The discoveries so far have been limited to coplanar systems (within $\sim 4^{\circ}$), so one might question the relevance of K-L cycles in this scenario. However, \citet{martintriaud14} argued that this narrow distribution is the result of a strong detection bias imposed by the requirement of a consecutive transit sequence\footnote{Circumbinary planets that are misaligned by $\gtrsim10^{\circ}$ frequently miss transits, creating a sparse transit sequence which is harder to identify.}. There are also theoretical arguments for the existence of misaligned circumbinary planets, as a result of planet-planet scattering (e.g. \citealt{chatterjee08} in the context of single stars), a misaligned disc (e.g. 99 Herculis, \citealt{kennedy12}, KH 15D, \citealt{winn04}) or an interaction with an outer tertiary star \citep{munoz15,martin15,hamers15}.

%Whilst circumbinary planets have been long known within a niche of authors (e.g. \citealt{rudaux37,lem61,lucas79}) and astronomers (e.g. \citealt{borucki84,dvorak86,schneider94}), 

In a circumbinary planetary system the outer orbital angular momentum becomes vanishingly insignificant with a decreasing outer mass. This is qualitatively different to the other systems in Fig.~\ref{fig:zoo_diagram} and it means that the classic studies of \citet{lidov62} and \citet{kozai62} are no-longer entirely applicable. Furthermore, since the K-L effect only modulates the eccentricity of the inner orbit, not the outer, it is not possible for an inner binary to induce a K-L cycle on a circumbinary planet. \citet{naoz13b} and \citet{liu15} derived equations for the amplitude and limits of the K-L effect for three arbitrary mass bodies, but did not apply them to circumbinary planets. {\it We are motivated to demonstrate the clear and significant difference between circumbinary planets and other three-body systems, and to clear up any possible misconceptions.}

The plan of the paper is as follows: In Sect.~\ref{sec:theory} we present equations for K-L theory, analyse the different limiting cases and consider the competing secular effect of general relativistic precession, which may act to suppress K-L cycles. We then apply the theory in Sect.~\ref{sec:application} to determine what parameter space of circumbinary planets may be able to induce K-L cycles before concluding in Sect.~\ref{sec:conclusion}.

\section{Theory}
\label{sec:theory}

\subsection{Equations for the induced eccentricity and critical mutual inclination}	
\label{sec:equations}

Consider a hierarchical triple system like those shown in Fig.~\ref{fig:zoo_diagram}, which we model as two Keplerian binary orbits. The three bodies have masses $M_1$ (navy blue in Fig.~\ref{fig:zoo_diagram}), $M_2$ (blue) and $M_3$ (light blue). Both orbits are defined by osculating orbital elements for the period, $P$, semi-major axis, $a$, eccentricity, $e$, argument of periapse, $\omega$, inclination $I$ and longitude of the ascending node, $\Omega$. We use the subscripts ``in" and ``out" to denote each orbit. The mutual inclination between the two orbits, $\Delta I$, is calculated by

\begin{equation}
\label{eq:DeltaI}
\cos\Delta I = \sin I_{\rm in}\sin I_{\rm out}\cos\Delta \Omega + \cos I_{\rm in}\cos I_{\rm out},
\end{equation}
where $\Delta \Omega = \Omega_{\rm in} - \Omega_{\rm out}$. The orbital angular momenta of the inner binary and outer binaries are 

\begin{equation}
\label{eq:angular_momentum_defintion}
G_{\rm in}=L_{\rm in}\sqrt{1-e_{\rm in}^2} \quad {\rm and} \quad G_{\rm out}=L_{\rm out}\sqrt{1-e_{\rm out}^2}
\end{equation}
where

\begin{equation}
\label{eq:binary_orbit}
L_{\rm in} = \frac{M_1M_2}{M_1+M_2}\sqrt{G(M_1+M_2)a_{\rm in}}
\end{equation}
and
\begin{equation}
\label{eq:binary_orbit}
L_{\rm out} = \frac{(M_1+M_2)M_3}{M_1+M_2+M_3}\sqrt{G(M_1+M_2+M_3)a_{\rm out}},
\end{equation}
and $G$ is the gravitational constant.

We analyse the three-body orbital evolution according to the quadrupole approximation of the Hamiltonian. In this case is no change to $a_{\rm in}$, $a_{\rm out}$ or $e_{\rm out}$ \footnote{ In the octupole level Hamiltonian there may be some variation in $e_{\rm out}$ but generally this is small. An interesting exception however was found by \citet{li14}, who showed that an eccentric inner binary ($e_{\rm in}\gtrsim 0.4$) may induce moderate eccentricity variations (up to $e_{\rm out}\sim 0.3$) on a massless outer body if it has a near-polar orbit.}. Under certain conditions there may be a significant variation in the $e_{\rm in}$ and $\Delta I$: a K-L cycle. Outside of K-L cycles $e_{\rm in}$ is constant. Additionally, both orbits will experience an apsidal and nodal precision (variations in $\omega$ and $\Omega$, respectively). 

The presence or absence of K-L cycles has significant implications for the stability of a three body system. For circular and coplanar triple systems there exists a rule of thumb that $a_{\rm out} \gtrsim 3 a_{\rm in}$ for stable orbits \citep{dvorak86,holman99,mardling01}. Around circular binaries, the stability limit generally moves inwards as $\Delta I$ increases and is only a weak function of the stellar mass ratio \citep{doolin11}. If the inner binary is eccentric, for example during a K-L cycle, then this stability limit is pushed outwards. An eccentric binary also makes the stability limit a more complex function of $\Delta I$ and the mass ratio \citep{doolin11,li14}. Seven of the known eight circumbinary systems have $e_{\rm in} \le 0.2$, with Kepler-34 being the only highly eccentric case ($e_{\rm in} = 0.5$, \citealt{welsh12}).

\citet{naoz13b} and \citet{liu15} derived a relation for the maximum inner eccentricity obtained, $e_{\rm in,max}$, as a function of the starting mutual inclination, $\Delta I_0$, the ratio of orbital angular momenta and the outer eccentricity,

\begin{align}
\label{eq:general_equation}
& 5\cos^2\Delta I_0 - 3 + \frac{L_{\rm in}}{L_{\rm out}}\frac{\cos\Delta I_0}{\sqrt{1-e_{\rm out}^2}} + \left(\frac{L_{\rm in}}{L_{\rm out}}\right)^2\frac{e_{\rm in,max}^4}{1-e_{\rm out}^2} \nonumber \\
& + e_{\rm in,max}^2 \left[3 + 4\frac{L_{\rm in}}{L_{\rm out}}\frac{\cos{\Delta I_0}}{\sqrt{1-e_{\rm out}^2}}  + \left(\frac{L_{\rm in}}{2L_{\rm out}}\right)^2\frac{1}{1-e_{\rm out}^2}\right] = 0,
\end{align}
where it is assumed that the inner binary is initially circular. Solutions for $e_{\rm in,max}$ only exist in a ``K-L active" region, which is bounded by lower and upper limits on the initial $\Delta I$, which we call $\Delta I_{\rm lower}$ and $\Delta I_{\rm upper}$. 

%\footnote{In \citet{naoz13b} it was also assumed that $e_{\rm out}=0$.}

The limiting mutual inclination for K-L cycles to occur is calculated by setting $e_{\rm in,max}=0$ in Eq.~\ref{eq:general_equation}, leaving a quadratic

\begin{figure*}  
\captionsetup[subfigure]{labelformat=empty}
\begin{center}  
	\begin{subfigure}[b]{0.99\textwidth}
		\includegraphics[width=\textwidth]{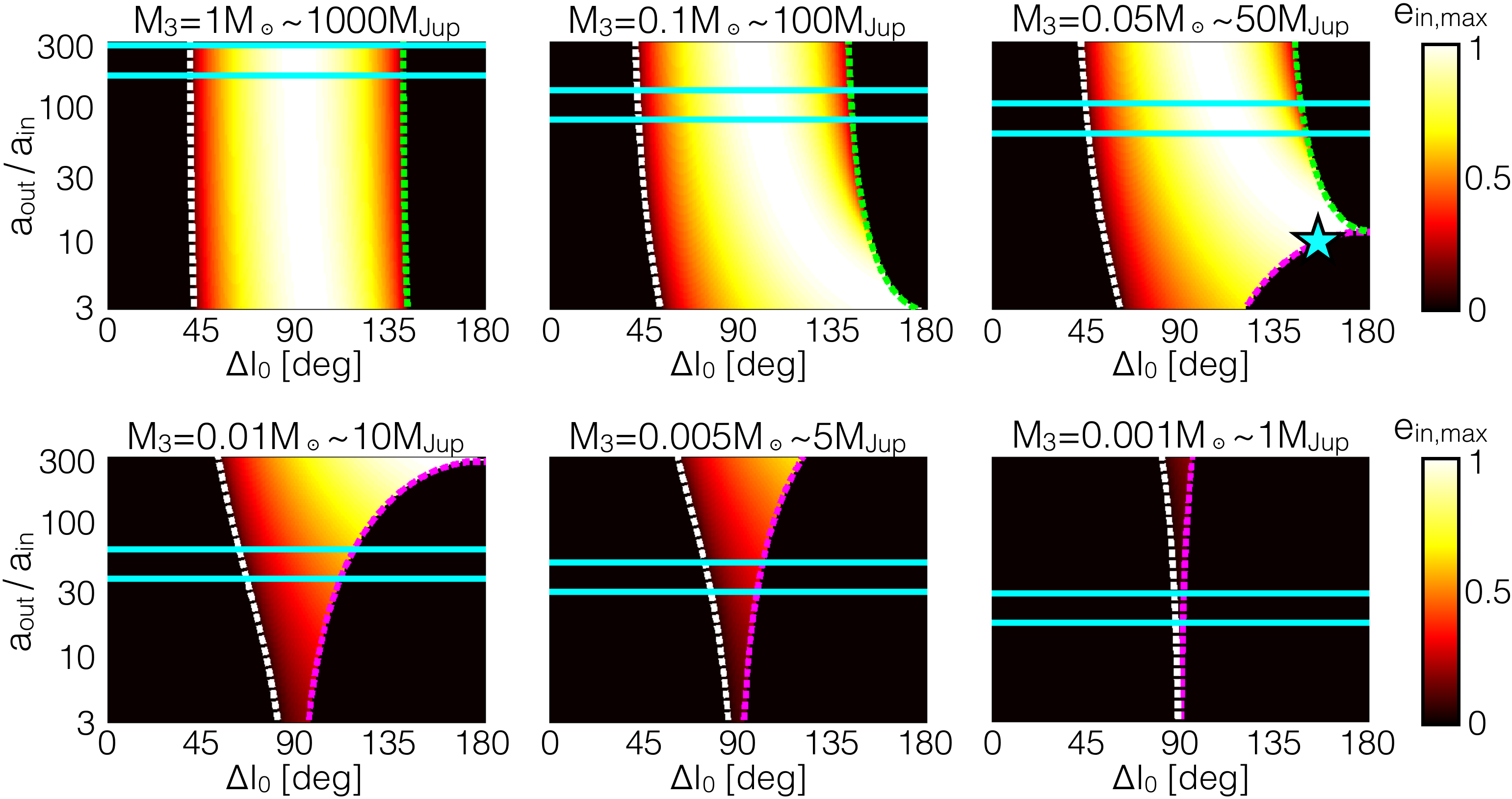} 
	\end{subfigure}
		
	\caption{Maximum inner eccentricity obtained during K-L cycles according to Eq.~\ref{eq:general_equation}, as a function of $a_{\rm out}/a_{\rm in}$ and $\Delta I_0$, for six different values of the outer mass $M_3$. The inner masses are set at $M_1=1M_{\odot}$ and $M_2=0.5M_{\odot}$. The initial eccentricities are both zero (and $e_{\rm out}$ is constant). The black regions correspond to where K-L cycles do not occur. These are separated from the K-L active regimes by a lower limit of $\Delta I_0$ (white dashed line on the left-hand limit from Eq.~\ref{eq:DeltaI_lower}) and an upper limit of $\Delta I_0$ (green and/or pink dashed lines on the right-hand limit from Eq.~\ref{eq:DeltaI_upper_1} and/or Eq.~\ref{eq:DeltaI_upper_2}, respectively). The light blue horizontal lines are upper limits on $a_{\rm out}/a_{\rm in}$ for K-L cycles to occur without being quenched by GR precession, for $a_{\rm in} = 0.0655$ AU ($P_{\rm in} = 5$ d) as the lower line and $a_{\rm in} = 0.3041$ AU ($P_{\rm in} = 50$ d) as the upper line. The light blue star symbol symbol in the top right subplot corresponds to the set of  N-body simulations in Fig.~\ref{fig:test_sharp}.
}\label{fig:kozai_sims}  
\end{center}  
\end{figure*}

\begin{equation}
\label{eq:DeltaI_quadratic}
5\cos^2\Delta I_{\rm lower,upper} - 3 + \frac{L_{\rm in}}{L_{\rm out}}\frac{\cos\Delta I_{\rm lower,upper}}{\sqrt{1-e_{\rm out}^2}} = 0.
\end{equation}
From solving the quadratic the lower limit of $\Delta I_0$ is defined as
\begin{align}
\label{eq:DeltaI_lower}
\begin{split}
   &\cos \Delta I_{\rm lower} = \frac{1}{10}\left[-\frac{L_{\rm in}}{L_{\rm out}}\frac{1}{\sqrt{1-e_{\rm out}^2}} + \sqrt{\left(\frac{L_{\rm in}}{L_{\rm out}}\right)^2\frac{1}{1-e_{\rm out}^2} + 60}\right].
\end{split}
\end{align}
The upper limit is simply the other solution to the quadratic,
\begin{align}
\label{eq:DeltaI_upper_1}
\begin{split}
   &\cos \Delta I_{\rm upper} = \frac{1}{10}\left[-\frac{L_{\rm in}}{L_{\rm out}}\frac{1}{\sqrt{1-e_{\rm out}^2}} - \sqrt{\left(\frac{L_{\rm in}}{L_{\rm out}}\right)^2\frac{1}{1-e_{\rm out}^2} + 60}\right] \\
   & {\rm for} \quad \frac{L_{\rm in}}{L_{\rm out}}\frac{1}{\sqrt{1-e_{\rm out}^2}} < 2,
\end{split}
\end{align}
but with the caveat that it is only defined when the outer orbit contains the majority of the angular momentum. Otherwise one must invoke a different upper limit, taken from \citet{lidov76}:
\begin{align}
\label{eq:DeltaI_upper_2}
\begin{split}
   &\cos \Delta I_{\rm upper} = -2\frac{L_{\rm out}}{L_{\rm in}}\sqrt{1-e_{\rm out}^2} \\
   & {\rm for} \quad \frac{L_{\rm in}}{L_{\rm out}}\frac{1}{\sqrt{1-e_{\rm out}^2}} \ge 2.
\end{split}
\end{align}
This second upper limit is not inherently respected by Eq.~\ref{eq:general_equation} when calculating $e_{\rm in,max}$ so it must be imposed manually.

\citet{liu15} compared numerical simulations with the analytic calculations using Eq.~\ref{eq:general_equation}, showing a very close match, but they only used Eq.~\ref{eq:DeltaI_upper_1} as an upper limit on $\Delta I_0$, not Eq.~\ref{eq:DeltaI_upper_2}. However in their tests the outer orbit always had a significant portion of the angular momentum, and hence Eq.~\ref{eq:DeltaI_upper_2} was never applicable.

\subsection{Limiting cases}
\label{sec:limiting_cases}

Kozai-Lidov cycles were originally derived in the limit of the {\it inner} restricted three-body problem, in which case $M_2 \rightarrow 0$ and hence $L_{\rm in}/L_{\rm out} \rightarrow 0$. In this case $\cos \Delta I_{\rm lower} \rightarrow \sqrt{3/5}$, and for the upper limit we use Eq.~\ref{eq:DeltaI_upper_1} to get $\cos \Delta I_{\rm upper} \rightarrow -\sqrt{3/5}$. This recovers the classic limiting mutual inclinations for K-L cycles: $\Delta I_{\rm lower} = 39.23^{\circ}$ and $\Delta I_{\rm upper} =140.79^{\circ}$. The calculation for $e_{\rm in,max}$ reduces from Eq.~\ref{eq:general_equation} to the simple well-known formula

\begin{equation}
\label{eq:emax_simple}
\left. e_{\rm in,max} \right |_{L_{\rm in}/L_{\rm out} = 0} = \sqrt{1-\frac{5}{3}\cos^2\Delta I_0},
\end{equation}
which is symmetric around $\Delta I_0=90^{\circ}$.

The opposite limiting case is the {\it outer} restricted three-body problem, where $M_3 \rightarrow 0$ , and hence $L_{\rm in}/L_{\rm out} \rightarrow \infty$. In this limit $\cos \Delta I_{\rm lower} \rightarrow 0$ and hence $\Delta I_{\rm lower} \rightarrow 90^{\circ}$. This means that the prograde K-L active region disappears. For the upper limit we use Eq.~\ref{eq:DeltaI_upper_2} to see that $\cos \Delta I_{\rm upper} \rightarrow 0$ and hence $\Delta I_{\rm upper} \rightarrow 90^{\circ}$, so the retrograde region disappears too. We recover the trivial limit of the outer restricted three-body problem, where there can be no K-L cycles since the binary feels no influence from the orbiting body, matching the work of \citet{lohinger03,farago10,doolin11}. The only dynamical effect on the circumbinary planet's orbit is an apsidal and nodal precession, at a rate calculated by \citet{farago10}.

\begin{figure}  
\begin{center}  
%	\begin{subfigure}[b]{0.99\textwidth}
%		\includegraphics[width=\textwidth]{test_smooth.png}  
%	\end{subfigure}
	\begin{subfigure}[b]{0.48\textwidth}
		\includegraphics[width=\textwidth]{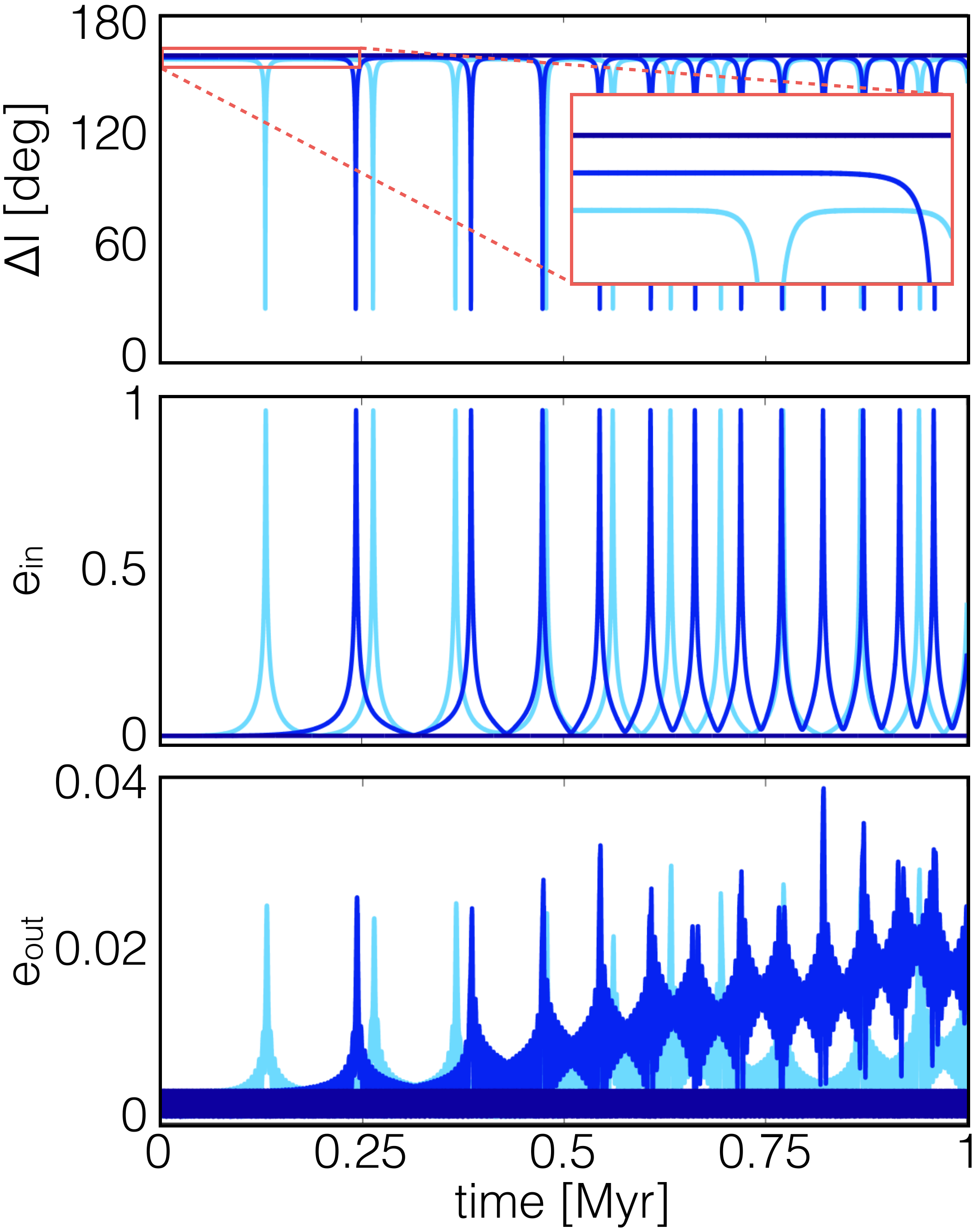}  
	\end{subfigure}	
\caption{Numerical simulations of K-L cycles, showing the evolution of $\Delta I$ (top panel), $e_{\rm in}$ (middle) and $e_{\rm out}$ (bottom). The parameters are $a_{\rm in}=$ 0.5 AU, $a_{\rm out}=$ 5 AU, $M_1=1M_{\odot}$, $M_2=0.5M_{\odot}$, $M_3=0.05M_{\odot}$ and $e_{\rm out}$ and $e_{\rm in}$ initially set at zero. Three different values of $\Delta I_0$ were tested: $157^{\circ}$ (light blue), $158^{\circ}$ (blue) and $159^{\circ}$ (navy blue). All other orbital angles were set to zero. The top panel includes a zoomed inset to show the small difference in starting conditions. The numerical values of $e_{\rm in,max}$ are 0.9894 ($\Delta I_0=157^{\circ}$), 0.9899 ($\Delta I_0=158^{\circ}$) and $1.63\times10^{-4}$ ($\Delta I_0=159^{\circ}$), and hence in the last case there is no K-L cycle. The analytic predictions for $e_{\rm in,max}$ are correct to within $0.04\%$ error.
}\label{fig:test_sharp}  
\end{center}  
\end{figure}

\subsection{Competing secular timescales}
\label{sec:timescales}

Even if a circumbinary system is said to be in a K-L active regime according to Sect.~\ref{sec:theory}, it is still possible that K-L cycles will not occur. This is because general relativity (GR) induces a competing precession on the inner binary, and out of the two effects the slower one will be suppressed\footnote{Tidal and rotational bulges also induce a precession on the binary's orbit but these effects are generally only significant for very tight binaries.}. The K-L timescale is

\begin{equation}
\label{eq:kozai_timescale}
\tau_{\rm K-L} = \frac{4}{3}\left(\frac{a_{\rm out}^2}{a_{\rm in}}\right)^{3/2}\sqrt{\frac{M_1+M_2}{GM_3^2}},
\end{equation}
which is taken from \citet{fabrycky07} but has been converted to be a function of semi-major axes. The GR timescale is

\begin{equation}
\label{eq:GR_timescale}
\tau_{\rm GR} = \frac{2\pi}{3}\frac{a_{\rm in}^{5/2}c^2}{G^{3/2}(M_1+M_2)^{3/2}},
\end{equation}
where $c$ is the speed of light \citep{fabrycky10}. 

An approximate limit for the suppression of K-L cycles is

\begin{equation}
\label{eq:criterion_simple}
\left. \frac{\tau_{\rm K-L}}{\tau_{\rm GR}}\right|_{\rm lim} \sim 1.
\end{equation}
This criterion assumes a sharp cut-off of K-L cycles, whereas in reality there is likely a smooth transition between a K-L dominated regime and a GR dominated regime. We are also ignoring the possibility of resonances between the two secular effects, which may in fact boost the modulation of $e_{\rm in}$ \citep{naoz13a}. Nevertheless, this criterion is deemed sufficient for the purposes of the simple analysis presented in this paper. Evaluating Eq.~\ref{eq:criterion_simple} using Eqs.~\ref{eq:kozai_timescale} and \ref{eq:GR_timescale} yields

\begin{equation}
\label{eq:criterion_full}
\left. \frac{a_{\rm out}}{a_{\rm in}}\right|_{\rm lim} \sim \left[a_{\rm in}\frac{M_3}{(M_1+M_2)^2}\frac{c^2\pi}{2G} \right]^{1/3}.
\end{equation}
This provides an upper limit on the tightness of circumbinary systems for which K-L cycles may be excited, lest they be quenched by general relativistic precession.

\section{Parameter space exploration}
\label{sec:application}

We explore how the Kozai-Lidov effect behaves in an example triple system consisting of an inner stellar binary with $M_1=1M_{\odot}$ and $M_2=0.5M_{\odot}$ and an outer body with $M_3$ ranging from a star down to a planet. Different values of $M_1$ and $M_2$ would change the limits and amplitude of the K-L effect, but in our example we use the mean stellar masses of the known circumbinary systems from {\it Kepler}, in order to make it as representative as possible. From Sect.~\ref{sec:equations} the amplitude and limits of K-L cycles are functions of the ratio of angular momentum, and consequently in our tests we only need to probe different values of the ratio $a_{\rm out}/a_{\rm in}$, not the individual values. We start $e_{\rm in}$ and $e_{\rm out}$ at 0, but $e_{\rm in}$ may rise significantly during a K-L cycle.

In Fig.~\ref{fig:kozai_sims} we calculate $e_{\rm in,max}$ from Eq.~\ref{eq:general_equation} as a function of $a_{\rm out}/a_{\rm in}$ and $\Delta I_0$, for six different outer body masses: $1M_{\odot}$, $0.1M_{\odot}$, $0.05M_{\odot}$, $0.01M_{\odot}$, $0.005M_{\odot}$ and $0.001M_{\odot}$. The smallest value of $a_{\rm out}/a_{\rm in}$ was 3, corresponding to the rough stability limit. The lower limit (left-hand limit) of the K-L active region is demarcated in a white dashed curve (Eq.~\ref{eq:DeltaI_lower}). The two different upper limits (right-hand limits), chosen according to the value of $M_3$ and $a_{\rm out}/a_{\rm in}$, are depicted as green (Eq.~\ref{eq:DeltaI_upper_1}) and pink (Eq.~\ref{eq:DeltaI_upper_2}) dashed curves.

The relative timescales of K-L cycles and GR precession set an upper limit in $a_{\rm out}/a_{\rm in}$ on the K-L active region, as a function of $a_{\rm in}$ (Eq.~\ref{eq:criterion_full}). In Fig.~\ref{fig:kozai_sims} two light blue horizontal lines denote the limiting ratio $a_{\rm out}/a_{\rm in}$ for $a_{\rm in} = 0.0655$ AU ($P_{\rm in} = 5$ d, lower line) and $a_{\rm in} = 0.3041$ AU ($P_{\rm in} = 50$ d, upper line). These two example values of $a_{\rm in}$ are not completely arbitrary as they cover the range of binaries known to host circumbinary planets \citep{welsh14}.

For $M_3=1M_{\odot}$ we have a triple star system like in Fig.~\ref{fig:zoo_diagram}a, which is well-modelled by a classic K-L regime with limits at $39^{\circ}$ and $141^{\circ}$. Here $e_{\rm in,max}$ is solely a function of $\Delta I_0$ and is well-approximated by Eq.~\ref{eq:emax_simple}. The GR timescale only impinges upon the widest systems within this parameter space. As $M_3$ decreases to $0.1M_{\odot}$, an M-dwarf, the K-L active regime is relatively unchanged for wide systems but shifts towards the retrograde region for tight systems. This asymmetry was noted by \citet{liu15}.

Within the brown-dwarf regime ($M_3=0.05M_{\odot}$) there is a significant change in the K-L behaviour for tight triple systems ($a_{\rm out}/a_{\rm in}< 15$). The high-amplitude eccentricity modulations ($e_{\rm in,max}\gtrsim 0.9$) are confined to retrograde orbits. Also the second upper limit (Eq.~\ref{eq:DeltaI_upper_2}) becomes applicable, causing the upper limit of the K-L active region to ``turn around" and move back towards $90^{\circ}$. 

%\footnote{It is a funny coincidence that a significant change in the K-L effect coincides with the transition between stars and planets.}

Within the planetary regime between roughly 1 and 10 $M_{\rm Jup}$, we see that K-L cycles become restricted to a narrow slither of angles either side of a polar $\Delta I_0=90^{\circ}$ orbit. The amplitude also decreases as the outer body becomes smaller. Furthermore the K-L timescale is also so slow that GR precession suppresses it in all but the closest systems. For reference, the observed circumbinary planets have been found predominantly near critical stability limit ($a_{\rm out}\sim 3a_{\rm in}$), with low eccentricities ($e_{\rm out}<0.2$), have mass $M_3<1M_{\rm Jup}$ and are nearly coplanar ($\Delta I \lesssim 4^{\circ}$) \citep{welsh14}.

%The prevalence of K-L cycles in near-polar orbits  may appear to contradict the results of \citet{li14}, who found that polar circumbinary orbits were actually more stable than other mutual inclinations, i.e. the inner stability limit resides closer to the binary. However \citet{li14} analysed massless circumbinary planets, for which K-L cycles are impossible. Our results for $M_3 = 1M_{\rm Jup}$ show that whilst K-L cycles at near-polar angles do persist down to this small mass, they have a very small amplitude: $e_{\rm in,max} \sim 0.03$. The work of \cite{li14} is therefore likely still applicable as long as $M_3 \lesssim 1 M_{\rm Jup}$. 

One interesting property is that the upper limit of K-L cycles calculated by Eq.~\ref{eq:DeltaI_upper_2} (pink dashed curve in Fig.~\ref{fig:kozai_sims}) imposes a sharp transition between K-L active and inactive regions. As an example, for $M_3=0.05M_{\odot}$ and $a_{\rm out}/a_{\rm in}=10$ there is a sharp transition at $\Delta I_0=159^{\circ}$ (marked with a light blue star symbol in the top right subplot in Fig.~\ref{fig:kozai_sims}). We demonstrate this with an N-body numerical solution of the exact equations of motion \footnote{ REBOUND (\citealt{rein12}, http://github.com/hannorein/rebound), with a 14-15th-order adaptive time step integrator \citep{rein15}.} in Fig.~\ref{fig:test_sharp}, showing the time-evolution of the osculating orbital elements $\Delta I$, $e_{\rm in}$ and $e_{\rm out}$ for three different values of $\Delta I_0$: $157^{\circ}$ (light blue), $158^{\circ}$ (blue) and $159^{\circ}$ (navy blue). The N-body code does not include GR, which is reasonable since the K-L timescale is significantly faster in this example. There is a sharp transition between K-L active and inactive regions at $\Delta I_0=159^{\circ}$. The maximum numerical value $e_{\rm in}$ matches the analytic prediction to within $0.04\%$ error.

Figure.~\ref{fig:test_sharp} shows that $e_{\rm out}$ remains close to zero but not exactly. For binaries undergoing a K-L cycle the outer eccentricity experiences low-amplitude periodic rises, in phase with the K-L cycle of the inner binary. This is not a K-L cycle being induced on the outer body itself, but rather a response to the high-amplitude eccentricity modulation of the orbited binary. 

Relaxing the requirement of a circular outer orbit moves the sharp upper limit towards $90^{\circ}$. In the Fig.~\ref{fig:test_sharp} example with $e_{\rm out}=0.5$ there is a shift in $\Delta I_{\rm upper}$ from $159^{\circ}$ to $144^{\circ}$, according to Eq.~\ref{eq:DeltaI_upper_2}, which we verified in an N-body simulation. Like before, the simulation showed little variation in $e_{\rm out}$ over time.

%
%\begin{table}
%\caption{Analytic and numerical values of $e_{\rm 1,max}$ near the lower and upper limits of $\Delta I_0$} %title of the table
%\centering % centering table
%\begin{tabular}{| c | c | c | c |} % creating eight columns
%\hline\hline %inserting double-line
%$\Delta I_0$ [deg] & analytic & numerical & error \%\\ 
%[0.5ex] 
%\hline % inserts single-line
%53 & 0 & $5.41\times10^{-4}$ & 0.05\\ 
%54 & 0.0375 & 0.0301& 19.74\\
%55 & 0.1174 & 0.1154 & 1.70\\
%\hline % inserts single-line
%157 & 0.9894 & 0.9894 & 0.00\\
%158 & 0.9903 & 0.9899 & 0.04\\
%159 & 0 & $1.63\times10^{-4}$ &  0.02\\
%\hline % inserts single-line
%
%\end{tabular}
%\label{tab:sim_comparison}
%\end{table}

%\begin{table}
%\caption{Analytic and numerical values of $e_{\rm 1,max}$ near the upper limit of $\Delta I_0$ for the simulations in Fig.~\ref{fig:test_sharp}} %title of the table
%\centering % centering table
%\begin{tabular}{| c | c | c | c |} % creating eight columns
%\hline\hline %inserting double-line
%$\Delta I_0$ [deg] & $e_{\rm in,max}$ (analytic) & $e_{\rm in,max}$ (numerical) & error \%\\ 
%[0.5ex] 
%\hline % inserts single-line
%157 & 0.9894 & 0.9894 & 0.00\\
%158 & 0.9903 & 0.9899 & 0.04\\
%159 & 0 & $1.63\times10^{-4}$ &  0.02\\
%\hline % inserts single-line
%
%\end{tabular}
%\label{tab:sim_comparison}
%\end{table}

\section{Conclusion}
\label{sec:conclusion}

The near-complete absence of potentially destabilising Kozai-Lidov cycles in the context of circumbinary planets has positive ramifications for the existence of misaligned systems. It frees us from the narrow confines of coplanarity and broadens the parameter space for potential discoveries. Based on studies of the abundance of circumbinary planets, if a presently-hidden population of misaligned systems were to exist, this would imply that planets are surprisingly more abundant around two stars than one \citep{armstrong14,martintriaud14}. 

There are also theoretical implications in the context of close binary formation by a combination of K-L cycles and tidal friction. Circumbinary planets cannot activate this mechanism. Circumbinary brown dwarfs may but only in close, retrograde orbits.

Observational evidence of such misaligned systems is presently lacking, but fortunately there exists several methods for their discovery: eclipse timing variations \citep{borkovits11}, astrometry \citep{sahlmann15} and sparse transits on both eclipsing and non-eclipsing binaries \citep{martintriaud14,martintriaud15}.

\section*{Acknowledgements}
We thank Smadar Naoz for her thorough reading of the manuscript and insightful comments. We also greatly benefited from discussions with Dan Fabrycky, Dong Lai, Rosemary Mardling and St{\'e}phane Udry. Our paper was aided by the helpful comments of an anonymous referee. We finally thank the teams behind \href{http://adsabs.harvard.edu}{ADS} and \href{http://arxiv.org/archive/astro-ph}{arXiv}.

\end{document}